# Novel magnetic and ferroelectric behaviors observed in alpha $Fe_2O_3$ particles*

Zhi Ma(马治)[1,2]**, Long Zhou(周龙)[1], Xu-Long Zhang(张旭龙)[1], Hong-Fei Liu(刘红飞)[1], Huan-Ming Chen(陈焕铭)[1], Fu Zheng(郑富)[1], Hua Gao(高华)[1]

[1] *School of Physics and Electronic-Electrical Engineering, Ningxia University, Yinchuan 750021, China*

[2] *State Key Laboratory of High-Efficiency Utilization of Coal and Green Chemical Engineering, Ningxia University, Yinchuan 750021, China*

*Supported by the Natural Science Foundation of Ningxia under Grant No. 2018AAC03060

**Corresponding author. Email :mazhi@nxu.edu.cn

%Email: mazhi@nxu.edu.cn; zhlnxu@163.com; zhxlnxu@163.com ； lhf0078@126.com; bschm@163.com; zhengfu-x@163.com; physmater@163.com；

%Tel.   13619587231

## Abstract

Alpha $Fe_2O_3$ powders have been prepared by the reduction reaction method with $NaHB_4$ as reducing agent and followed a conventional sintering process. The XRD pattern with Rietveld refinement profile reveal that the prepared $Fe_2O_3$ with corundum structure (hematite). VSM loop exhibits obvious room-temperature weak ferromagnetism, a pinched hysteresis loop may introduced by the shape anisotropy effect. The simultaneous ferroelectric behavior of α-$Fe_2O_3$ with "five-fold" ferroelectric hysteresis loops approves that this structured $Fe_2O_3$ can be known as a novel multiferroic material.

**PACS:**   51.60.+a, 77.84.-s, 77.80.Dj

Transition metal oxides material is one of the fastest growing field during the past several decades. Interest in this field is attributable to the increasing numbers of its practical applications, such as recording material, solar cell, pigments, microwave absorption, catalysts, gas sensors, environmental pollutant clean up agents, electrode materials, biological materials and clinical diagnosis and treatment, etc. Meanwhile, physics of transition metal oxides has always been exciting for understanding the basic magnetism/ferroelectricity and their potential use in diverse applications.

Among the magnetic transition metal oxides, the iron oxide are widely studied in recent literature owing to its specific properties such as bio-compatibility and non-toxicity, as well as their fascinating structural and magnetic properties. Apart from the amorphous $Fe_2O_3$, six polymorphs have been identified so far, α-$Fe_2O_3$ (hematite) with corundum structure, β-$Fe_2O_3$ (barite) with bixbyite structure, γ-$Fe_2O_3$ (maghemite) with cubic crystal structure of the inverse spinel type, the orthorhombic ε-$Fe_2O_3$, rhombohedral phase η-$Fe_2O_3$ and monoclinic phase ζ-$Fe_2O_3$ [1-5]. Among them, hematite (α-$Fe_2O_3$) is the most thermodynamically stable phase. As to the hematite, it has three critical temperatures: Neel temperature ($T_N$, about 950 K), blocking temperature ($T_B$, about 50 K), and the Morin temperature of $T_M$ ~260 K [6,7]. It is known that all the three ordering temperatures for α-$Fe_2O_3$ depend on their size, shape, microstructure, the strain and the defects[8].

To date, hematite (α-$Fe_2O_3$) shows remarkable physical properties, both for fundamental investigations and practical applications. However, to the knowledge of the authors, the related research on the combination of magnetical and ferroelectrical behaviors has not been reported thus far in literature. In this work, the study is accompanied by a thorough investigation of the magnetic and ferroelectric properties of the hematite particles, revealing some interesting anomalies. The results of this research can provide further insight from many perspectives-not only to satisfy desire for fundamental knowledge but also due to prospective applicability.

A conventional reduction reaction method was used to prepare the precursor of hematite. All reagents were of analytical grade purity and used as received without further purification. Analytical reagent $Fe(NO_3)_3·9H_2O$ (0.05 M) was added to 200 ml deionized water and vigorously stirred for 2h. Then, 200 ml aqueous solution of sodium borohydride ($NaHB_4$) (0.4 M) was added dropwise to the above transparent solution to make the reaction complete. Black powders were separated and washed with deionized water several times. Finally, the obtained powders were dried and calcined at 900 °C for 4h. The calcined powders were mixed thoroughly with a polyvinyl alcohol (PVA) binder solution, and then pressed into disk-shaped pellets of 10 mm in diameter and about 2 mm in thickness. After that, the disk samples were sintered at the temperatures of 900 °C for 1h in an alumina crucible in air atmosphere. Ag electrodes were prepared on both sides of the samples by firing at 500 °C for 1h in the air for measuring ferroelectric properties.

The X-ray diffraction (XRD) patterns were obtained on a SmartLab 9KW-XRD diffractometer (Cu K$_α$, λ = 0.154059 nm, Rikagu, Japan) to investigate the crystalline phases of the products. XRD data was collected in the range of 20° to 85° with 0.01° step and the X-ray was generated at 45 kV and 200 mA. A scanning electron microscope (SEM, SNE-4500, SEC Co., South Korea) was employed to investigate the morphologies. Magnetic properties were studied by vibrating sample magnetometer (VSM, MicroSense-EZ9, USA) with a maximum magnetic field of 2.6 T. The electric field induced polarization were measured by a precision ferroelectric analyzer (Premier II, Radiant Technologies, USA).

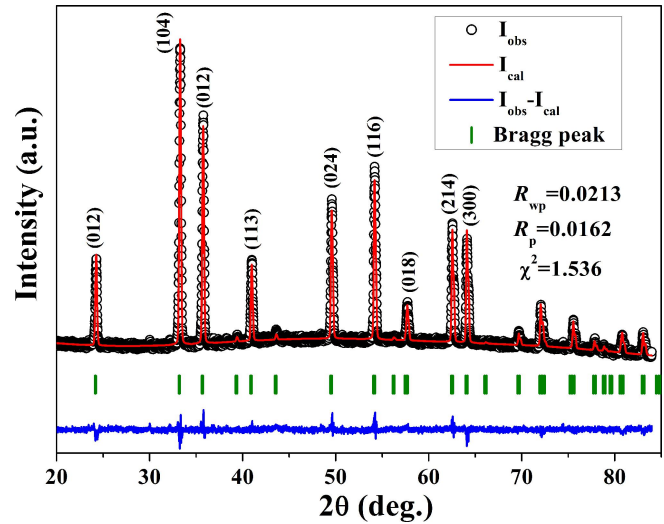

**Fig. 1.** The X-ray pattern was refined using GSAS-EXPGUI program based on the Rietveld method, the Miller indices of the Bragg peaks are also shown. The Rietveld profile fitting parameters are: $R_{wp}$= 0.0213, $R_p$ = 0.0162, $R_e$ = 0.0139, $χ^2$=1.536.

XRD was used to investigate the phase structure of the powders. In Fig. 1, XRD pattern is shown along with Rietveld refinement profile and the difference curve. The GSAS-EXPGUI program has been used to perform the Rietveld refinement process[9]. It can be seen from Fig. 1 that the reliability factors $R_{wp}$ and $R_p$ are 2.13% and 1.62%, respectively, indicating that the observed pattern is successfully fitted by calculated one. Here, $R_{wp}$ designated as a good fitting parameter is smaller than 10% indicating that the fitting convergence is attained. The very small value of $R_e$ and $\chi^2>1$ confirm that errors are no longer dominated by counting statistics and a fully refined structure is achieved[10]. The estimated crystal parameters are a=b=0.5037176 nm, c=1.375353 nm, which is close to the report in literature[7].

The Miller indices for the peaks are also shown and this confirms that the sample has hematite type (α-$Fe_2O_3$) structure in the space group R-3c(167) in hexagonal setting, since $Fe_2O_3$ is polymorphic material having six crystal arrangements of $Fe^{3+}$ and $O^{2-}$ ions resulting in α, β, γ, ε, η and ζ-$Fe_2O_3$ phases[4,11-14]. Out of all the phases, hematite (α-$Fe_2O_3$) is the most stable phase of iron oxide at ambient conditions. The XRD pattern also show that, no peaks from other phases are found, samples exhibit diffraction patterns of single phase well-crystalline α-$Fe_2O_3$, according to the standard diffraction spectrum of α-$Fe_2O_3$ (JCPDS No. 33-0664). It indicates that the sample successfully prepared during sintering process and no additional reaction occurred.

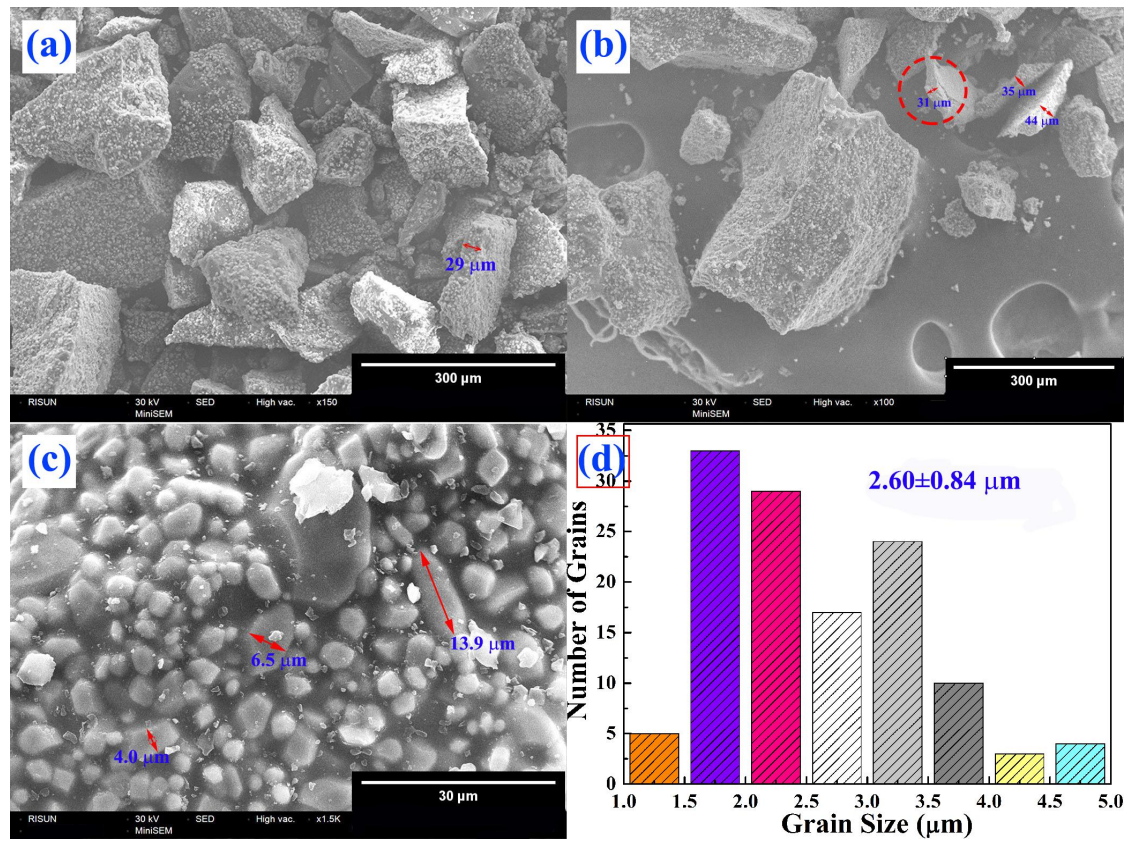

**Fig. 2.** (a) Scanning electron microscopy image of α-$Fe_2O_3$ ceramic sintered at 900 °C and (b) low-magnification and (c) high-magnification images of the ceramic powders and (d) grain size distributions of the particles.

Fig. 2 displays the SEM photographs of the α-$Fe_2O_3$ sample after densification at 900 ºC, the average particle size distribution diagrams are also shown in the pictures. As observed, the sample shows very fine agglomerated particles without a definite shape or it with irregular well-defined rock like grains (Fig. 2a). It can be found that the sintered ceramics have typical polyhedron structures with diameter sizes ranging from dozens to several hundred micrometers. Some grains exhibit plate-shaped structure.

Fig. 2b shows low-magnification images of the α-$Fe_2O_3$ powders, where several plate-shaped grains with thickness about 30 to 40 micrometers could been observed. It can be seen from Fig. 2c that the angular-shaped and big polyhedrons are made up of small particles. Fig. 2d gives the particle size distribution from 1.2 to 4.95 μm with 2.60 μm as average grain size. It can be inferred that the larger grain structure in sintered powders is reconstructed from the agglomeration of smaller particles. The results revealed the existence of no single particle size distribution and evidence of significant particle agglomeration.

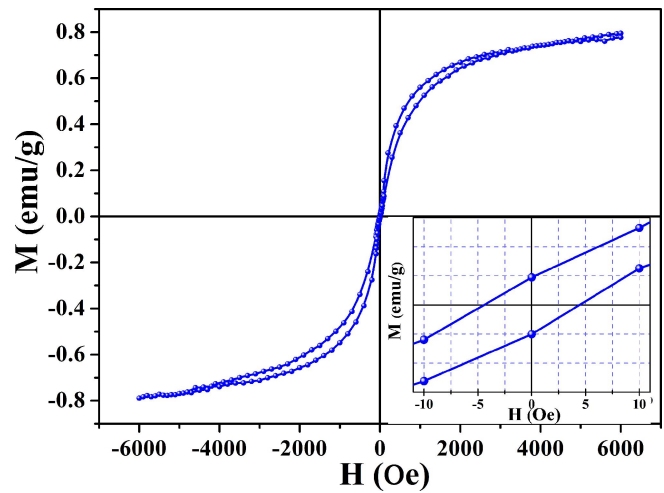

**Fig. 3.** Hysteresis loop of the α-$Fe_2O_3$ powder measured at room temperature.

The magnetic properties of the sample are measured by VSM at room temperature. The hysteresis loop obtained is shown in Fig. 3. It can be seen from the hysteresis loop that the saturation magnetization (Ms), the remnant magnetization (Mr) and the coercivity (Hc) of the α-$Fe_2O_3$ powder are 0.7 emu/g, 0.005 emu/g and 4.0 Oe, respectively. The value of saturation magnetization is much lower than value of maghemite (γ-$Fe_2O_3$) nanoparticles (30 emu/g)[15]. Meanwhile, M-H curve of α-$Fe_2O_3$ powder indicates a pinched hysteresis loop with finite but small remnant polarization. This may be due to the plate shape of the microparticles (Fig. 2b) most probably introduced a shape anisotropy effect and additionally decreased overall coercivity of the sample[16].

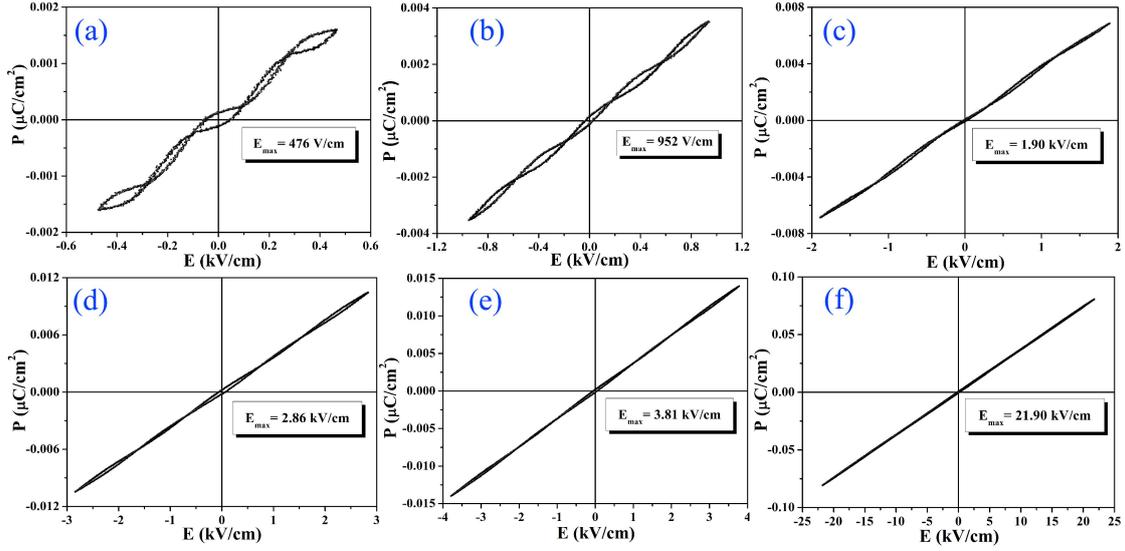

**Fig. 4.** Polarization-electric field hysteresis loops of α-$Fe_2O_3$ at different electric fields and 10 Hz.

Fig. 4 shows ambient temperature polarization-electric(P-E) hysteresis loops of α-$Fe_2O_3$ with an $E_{max}$ increasing from 476 V/cm to 21.90 kV/cm. It is interesting to see that the P-E curve shows a pinched shape at low electric field. As the rising of electric field, the pinched loop becomes slimmer and slimmer. When the electric field is higher than 3.81 kV/cm, the distortion in hysteresis loops disappears and the hysteresis loop becomes a straight line. The most interesting point of the figure is that "five-fold" hysteresis loops was observed when the electric field are 476 V/cm and 952 V/cm. This phenomena is anomalous to the reports in literature.

Various physical mechanisms have been proposed to explain the observed pinched hysteresis loops. A complete E-T phase diagram and exemplary isothermal dependence of polarization on electric fields have been developed by Fesenko et al.[17]. In which, 1, 2, 3, 4 and 6 fold hysteresis loops were observed and in the phase-stability regions of the diagram, the multi-fold hysteresis loops usually corresponding to the field-induced structure phase transitions. Li Jin et al. assumed that those reversible antiferroelectric-to-ferroelectric phase transitions lead to obvious pinched hysteresis loops[18]. More specifically, the appearance of pinched hysteresis loops should be attributed to that a metastable ferroelectric phase coexist with the antiferroelectric phase in the poled samples and the induced ferroelectric phase will return to the antiferroelectric phase after removing the electric field[19]. Note that the pinched hysteresis loops were also explained in terms of the reversible "non-polar" phase to ferroelectric phase transitions[20]. However, K. Carl and K.H. Hardtl attributed the pinched hysteresis loops to an internal bias field, and the volume effect, domain effect and grain boundary effect may account for the occurrence of such an internal bias[21]. Zhang et al. assumed that the pinched loops results from the pinning effect of the oxygen vacancies induced defect dipoles[22]. Very recently, the researchers suggested that the pinched shape was probably related to strong interactions between defect dipoles and the polarization[23]. To understand its microscopic mechanism of the "five-fold" hysteresis loops, further physics research and exploring are necessary.

In conclusion, $\alpha$-$Fe_2O_3$ powders are prepared using a conventional reduction reaction method. Novel room-temperature ferromagnetism and ferroelectrics can be observed in the structured $\alpha$-$Fe_2O_3$ particles. A pinched magnetic hysteresis loop reveal that the plate shape of the microparticles most probably introduced a shape anisotropy effect and additionally decreased overall coercivity of the sample. When the applied electric field are 476 V/cm and 952 V/cm, "Five-fold" ferroelectric hysteresis loops could be observed. But the microscopic mechanism should been further researched.

## References

[1] S. Pan, W. Huang, Y. Li, L.L. Yu, R.J. Liu, Mater. Lett. 262 (2020) 127071.

[2] P.B. Patil, S.B. Parit, P.P. Waifalkar, S.P. Patil, T.D. Dongale, Subasa C. Sahoo, P. Kollu, M.S. Nimbalkar, P.S. Patil, A.D. Chougale, Mater. Lett. 223 (2018) 178-181.

[3] D. B. Jiang, B. Y. Zhang, T. X. Zheng, Y. X. Zhang, X. Xu, Mater. Lett. 215 (2018) 23-26.

[4] J. TuCek, L. Machala, S. Ono, A. Namai, M. Yoshikiyo, K. lmoto, H. Tokoro, S. I. Ohkoshi and R. Zboril, Sci. Rep. 5 (2015) 15091.

[5] J. S. Lee, S. S. Im, C. W. Lee, J. H. Yu, Y. H. Choa and S. T. Oh, J. Nanopart. Res. 6 (2004) 627-631.

[6] N. Amin and S. Arajs, Phys. Rev. B 35 (1987) 4810-4811.

[7] A. S. Tej, P. Y. Koh, Prog. Cryst. Growth CH 55 (2009) 22-45.

[8] Y. Y. Xu, L. Wang, T. Wu, R. M. Wang, Rare Metals 38 (2019)14-19.

[9] B. H. Toby and R. B. Von Dreele, J. Appl. Cryst. 46 (2013) 544-549.

[10] L. B. McCusker, R. B. Von Dreele, D. E. Cox, D. Louër and P. Scardi, J. Appl. Cryst. 32 (1999) 36-50.

[11] D. A. Balaev, A. A. Dubrovskiy, S. S. Yakushkin, G. A. Bukhtiyarova, and O. N. Martyanov, Phys. Solid State 61 (2019) 345-349.

[12] J. S. Lee, S. S. Im, C. W. Lee, J. H. Yu, Y. H. Choa and S. T. Oh, J. Nanopart. Res.  6 (2004) 627-631.

[13] L. Kubickova, P. Brazda, M. Veverka, O. Kaman, V. Herynek, M. Vosmanska, P. Dvorak, K. Bernasek, J. Kohout, J. Magn. Magn. Mater. 480 (2019) 154-163.

[14] J. Y. Zhong, C. B. Cao, Y. Y. Liu, Y. N. Li and W. S. Khan, Chem. Commun. 46 (2010)

3869-3871.

[15] M. Hussain, R. Khan, Zulfiqar, T. Z. Khan, G. Khan, S. Khattak, M. U. Rahman, S. Ali, Z. lqbal, Burhanullah, K. Safeen, J. Mater. Sci-Mater. El. 30 (2019) 13698-13707.

[16] M. Tadic, D. Trpkov, L. Kopanja, S. Vojnovic, M. Panjan, J. Alloy. Compd. 792 (2019) 599-609.

[17] O. E. Fesenko, R. V. Kolesova and Yu. G. Sindeyev, Ferroelectrics 20 (1978) 177-178.

[18] L. Jin, J. Pang, R. Y. Jing, Y. Lan, L. Wang, F. Li, Q. Y. Hu, H. L. Du, D. Guo, X. Y. Wei, Z. Xu, L.Y. Zhang, G. Liu, J. Alloy. Compd. 788 (2019) 1182-1192.

[19] Y. P. Guo, M. Y. Gu, H. S. Luo, Y. Liu, and R. L. Withers, Phys. Rev. B 83 (2011) 054118.

[20] Y. P. Pu, M. T. Yao, H. R. Liu, T. Fromling, J. Eur. Ceram. Soc. 36 (2016) 2461-2468.

[21] K. Carl, K. H. Hardtl, Ferroelectrics 17 (1977) 473-486.

[22] H. B. Zhang, S. L. Jiang, and Y. K. Zeng, Appl. Phys. Lett. 93 (2008) 192901.

[23]S. Saremi, J. Kim, A. Ghosh, D. Meyers, and L. W. Martin, Phys. Rev. Lett. 123 (2019) 207602.